\documentclass[doublecol]{epl2}

\usepackage{epsfig}
\usepackage{amsmath,amssymb,wasysym,epsfig,capt-of,ifthen,calc}
\usepackage{bbm}
\usepackage{latexsym} 
\usepackage{setspace} 
\usepackage{array} 
\usepackage{delarray} 
\usepackage{afterpage} \usepackage{graphicx}
\usepackage{dcolumn}
\usepackage{bm}
\usepackage{float} \usepackage{supertabular} \usepackage{longtable}
\newcommand{\be}{\begin{equation}} \newcommand{\ee}{\end{equation}}
\newcommand{\bea}{\begin{eqnarray}} \newcommand{\eea}{\end{eqnarray}}

\usepackage{amsmath}\bibstyle{apsrev}

\title{Percolation Theory on Interdependent Networks\\ Based on Epidemic Spreading}
\shorttitle{Percolation Theory on Interdependent Networks} 

\author{Seung-Woo Son\inst{1} \and Golnoosh Bizhani\inst{1} \and Claire Christensen\inst{1}
\and Peter Grassberger\inst{1,2}\\ \and Maya Paczuski\inst{1}}
\shortauthor{Seung-Woo Son \etal}

\institute{
  \inst{1} Complexity Science Group, University of Calgary, Calgary T2N 1N4,
  Canada \\
  \inst{2} FZ J\"ulich, D-52425 J\"ulich, Germany, EU
}


\abstract{ We consider percolation on interdependent locally
treelike networks, recently introduced by Buldyrev {\it et al.},
Nature {\bf 464}, 1025 (2010), and demonstrate that the problem
can be simplified conceptually by deleting all references to
cascades of failures. Such cascades do exist, but their explicit
treatment just complicates the theory -- which is a
straightforward extension of the usual epidemic spreading theory
on a single network.   Our method has the added benefits that it
is directly formulated in terms of an order parameter and its
modular structure can be easily extended to other problems, {\it
e.g.} to any number of interdependent networks, or to networks
with dependency links. }

\begin{document}

\maketitle

On September 28, 2003, Italy experienced its most severe blackout
in over 20 years~\cite{Rosato2008}. The blackout's severity was
later attributed to the fact that an initial failure on the
physical power grid disrupted not only the grid, itself, but also
a computer network that depended on this grid for
electricity~\cite{Rosato2008,Buldyrev2010}. Since the grid's
substations were, in turn, dependent on this computer network for
their regulation, further failures in the grid ensued as
communication among the stations was lost. Ultimately, recursive
cascading failures throughout {\it both} networks occurred, and
both networks changed from percolating to
non-percolating~\cite{Rosato2008,Buldyrev2010,Parshani2010,Parshani2011}.
While this is a spectacular example of percolation on
interdependent networks, it is, by no means the only one:  in
fact, {\it most} real-world networks can be seen as having some
interdependency~\cite{Moreno2003,Motter2004,Kurant2006}. A clear
description of how perturbations propagate through such networks
-- {\it i.e.} of how perturbations can effect percolation
(fragmentation) -- is essential to understanding systems involving
interdependence, including economic markets, interrelated
technological and infrastructural systems, social networks,
disease dynamics, or human physiology.

Recent models of percolation on interdependent networks have been
described in terms of failures cascading back and forth between
the networks ~\cite{Buldyrev2010,Parshani2010,Parshani2011}. While
there is no doubt that percolation on interdependent networks {\it
can} be seen as a cascading phenomenon, the mathematics behind
such a description is cumbersome and far from transparent. Here we
simplify matters by omitting all aspects of cascading and treat
percolation on interdependent networks as an epidemic spreading
process in complete analogy to ordinary percolation. If one wants
to consider cascades explicitly, this can be done in a second
step, after the phase transition itself is well understood.  We
also show that percolation on dependency
networks~\cite{Parshani2011,Bashan2011}, which are single networks
composed of both connectivity links and dependency links, can be
described within the same paradigm.

Both the theory of~\cite{Buldyrev2010,Parshani2010,Parshani2011}
and the present paper deal only with {\it locally treelike} random
networks, for which mean field theory based on generating
functions becomes exact in the large system limit. The
cascade-based studies
in~\cite{Buldyrev2010,Parshani2010,Parshani2011} considered site
percolation networks from which a certain fraction $1-p$ of nodes
had been removed. On general networks (including lattices), this
site dilution can lead to a topological modification of the
networks~\cite{Son2011}, but in the present cases it just leads to
a trivial rescaling of the number of nodes and links. Thus we can
restrict ourselves, without loss of generality, to the case $p=1$
and only consider bond percolation, which further simplifies the
discussion.

In the following, we first recall the theory of epidemic spreading
(ordinary percolation theory) on single networks
\cite{Newman2005,Newman2001}, and then demonstrate how this can be
easily adjusted to accommodate percolation on interdependent
networks or on dependency networks. We only consider the limit of
large networks, where the number of nodes $N\to\infty$.

(i) \emph{Single network} - First consider a single (isolated)
random network with mean degree $z$, with an epidemic spreading
from some starting node. Let $S_i$ be the probability that node
$i$ is infected during this epidemic ({\it i.e.} that it is part
of the infinite percolating cluster). Its average over all nodes,
denoted as $S$, is taken as the order parameter of the model. The
probability that node $i$ is not infected is equal to the chance
that none of its neighbors are infected through their remaining
links: \be
    1-S_i = \prod_{\langle ij \rangle} (1-S'_j), \label{eq1}
\ee where the product runs over all neighbors of $i$ and $S'_j$ is
the probability that node $j$ is infected through a randomly
chosen edge {\it not} attached to $i$. When the graph is locally
treelike, all $S'_j$ are independent. Averaging Eq.~(\ref{eq1})
over all nodes gives then \be
     S = 1-\sum_{k} p(k) (1-S')^k \equiv 1-G_0 (1-S'), \label{eq2}
\ee where $p(k)$ is the probability that a node has $k$ links and
$G_0(x)$ is the corresponding generating function $G_0(x) \equiv
\sum_k p(k) x^k$. Similarly, one can write down the equation for
$S'$ \be
   S' = 1-\sum_{k} {k
      p(k) \over z} (1-S')^{k-1} \equiv 1-G_1 (1-S'), \label{eq3}
\ee where $z = G'_0(1)$ and $G_1(x) \equiv G'_0(x)/z$. For any
given degree distribution we can first solve Eq.~(\ref{eq3}) to
obtain $S'$, and then insert it into Eq.~(\ref{eq2}) to obtain
$S$. The percolation transition threshold $z_c$ is given by $S=0$
for $z<z_c$ and $S>0$ for $z>z_c$~\cite{Newman2005,Newman2001}.

\begin{figure}[t]
\center\includegraphics[width=0.8\columnwidth]{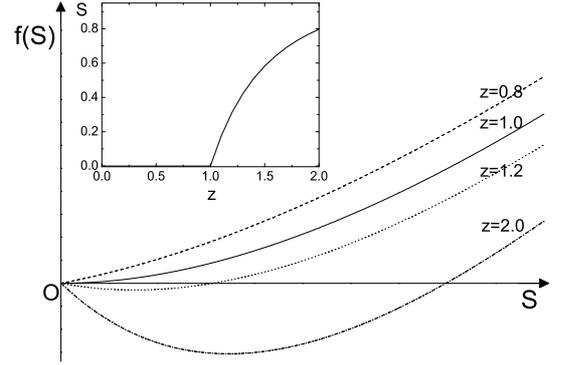}
\caption{Graphical solutions for ordinary percolation on a single
ER network. The inset shows the {\em continuous} change of order
parameter $S$ as $z$ increases. } \label{fig1}
\end{figure}

For instance, the degree distribution of Erd\"os-R\'enyi (ER)
graphs~\cite{Bollobas1985,Newman2001} is Poissonian. Therefore,
\bea
   G_0(1-S')
     &=& \sum_k {e^{-z} z^k \over k!} (1-S')^k = e^{-zS'} \nonumber \\
     &=& G_1(1-S'). \label{eq4}
\eea Thus, $S=S'$ and Eqs.~(\ref{eq2}) and (\ref{eq3}) give simply
$S=1- e^{-zS}$. Defining $f(S) = S - 1 + {e}^{-zS}$, one can find
the solution $S(z)$ by solving $f(S)=0$ graphically, as shown in
Fig.~\ref{fig1}. A continuous (`second order') phase transition is
clearly evident in the inset of Fig.~\ref{fig1}.

Equations~(\ref{eq2}) and (\ref{eq3}), are the order parameters
for a single network, where $S$ represents the probability that a
randomly chosen node places in the infinite percolating cluster
and $S'$ is the same probability, but defined when we pick an edge
randomly and look at the end node. These equations act as
fundamental `building blocks' or `modules' for treating
analogously the probability, on fully or partially interdependent
networks, to be connected to the infinite percolating clusters.

(ii) \emph{Two fully interdependent networks} -  Consider now two
networks $\cal{A}$ and $\cal{B}$,  where each node in ${\cal A}$
depends only on one node in ${\cal B}$ and {\it vice versa}. In
order for a node in network ${\cal A}$ to be part of the
percolating cluster, its partner in ${\cal B}$ must also be part
of that cluster.  Since this mapping is one-to-one we can merge
each node in ${\cal A}$ with its partner in ${\cal B}$ to have one
set of nodes, each with two sets of links.
 We define
$\cal{AB}$-clusters as subsets of nodes connected both in
$\cal{A}$ and in $\cal{B}$. More precisely a set of nodes $C =
\{i_1,i_2,\ldots i_m \}$ is an $\cal{AB}$-cluster if any two
points $i, j \in C$ are connected by two paths: one path using
only links $\in \cal{A}$ and nodes only $\in C$, and the other
using only links $\in \cal{B}$ and also using nodes only $\in C$.
We do not allow paths that involve nodes outside $C$, so
 $\cal{AB}$-clusters are \emph{self-sustaining}~\cite{Son2011}.

The probability that any node belongs to the infinite
$\cal{AB}$-cluster is equal to the probability to be linked to it
both via $\cal{A}$- and via $\cal{B}$-links.  That is, a node
looks out at its ${\cal A}$-links to see if it has a neighbor on
the percolating ${\cal AB}$-cluster. It also looks out via its
${\cal B}$ links. Only if it has a neighbor via both sets of links
is it a member of this cluster.  Therefore $S$ is simply a product
of the right hand side of Eq.~(\ref{eq2}) for networks ${\cal A}$
and ${\cal B}$, \be
     S = (1-G^{\cal{A}}_0(1-S'_{\cal{A}}))(1-G^{\cal{B}}_0(1-S'_{\cal{B}})),\label{eq5a}
     \ee
where the superscripts (subscripts) $\cal{A}$ and $\cal{B}$ refer
to networks $\cal{A}$ and $\cal{B}$ respectively and $S_{\cal{A}}
= S_{\cal{B}}=S$. Here, $S'_{\cal{A}}$ (and analogously
$S'_{\cal{B}}$) is defined as the probability that a node reached
by following a random $\cal{A}$-link is in the ${\cal
AB}$-cluster. For this to happen, its partner node -- which is a
random node from the point of view of network $\cal{B}$ -- has
also to be connected to the ${\cal AB}$-cluster via
$\cal{B}$-links. $S'_{\cal{A}}$ and $S'_{\cal{B}}$ are different
from each other since they depend on the degree distribution of
each network. When choosing edges at random, the end node of a
randomly chosen edge in network $\cal{A}$ belongs to the
$\cal{AB}$-cluster only when its partner node in network $\cal{B}$
is concurrently a member of this cluster, and {\it vice versa}.
The probabilities of these events occurring -- $S'_{\cal{A}}$ and
$S'_{\cal{B}}$ -- are given by \bea
     S'_{\cal{A}}&=& (1-G^{\cal{A}}_1(1-S'_{\cal{A}}))(1-G^{\cal{B}}_0(1-S'_{\cal{B}})), \nonumber \\
     S'_{\cal{B}}&=& (1-G^{\cal{B}}_1(1-S'_{\cal{B}}))(1-G^{\cal{A}}_0(1-S'_{\cal{A}})). \label{eq5}
\eea

\begin{figure}[t]
\center\includegraphics[width=0.85\columnwidth]{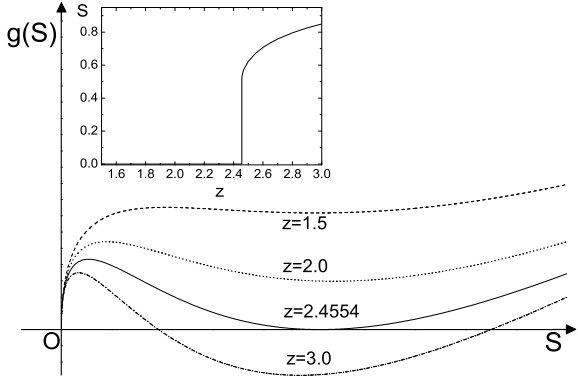}
\caption{Graphical solutions for the percolating cluster on two
fully coupled ER networks with the same mean degree $z$. The inset
 shows a {\em discontinuous} change of order parameter $S$ as
$z$ increases.} \label{fig2}
\end{figure}

Using this and Eq.~(\ref{eq4}) for two interdependent ER networks
with mean degrees $z_{\cal{A}}$ and $z_{\cal{B}}$ gives \be
    S = (1-e^{-z_{\cal A}S})(1-e^{-z_{\cal B}S}). \label{eq_ER}
\ee In particular, when $z_{\cal{A}} = z_{\cal{B}} = z$, the order
parameter obeys \be
     S = (1-e^{-zS})^2.     \label{eq_5}
\ee Defining $g(S) = S - (1 - e^{-zS})^2$, one can find
graphically the value of $S$ as a function of $z$ that solves
$g(S)=0$  (see Fig.~\ref{fig2}). This solution  shows a
discontinuous (`first order') phase transition (inset of
Fig.~\ref{fig2}), contrary to the result for the single network in
Fig.~\ref{fig1}. Demanding $g(S)=g'(S)=0$ we find a critical point
$S_c=0.511699\cdots$ and $z_c = 2.455407\cdots$. The value of
$z_c$ agrees with that given for ER networks
in~\cite{Buldyrev2010,Son2011} and our explicitly determined value
of the order parameter just above the critical point, $S_c$, can
also be obtained from appropriate combinations of their results.

\begin{figure}[t]
\centering
\includegraphics[width=0.72\columnwidth]{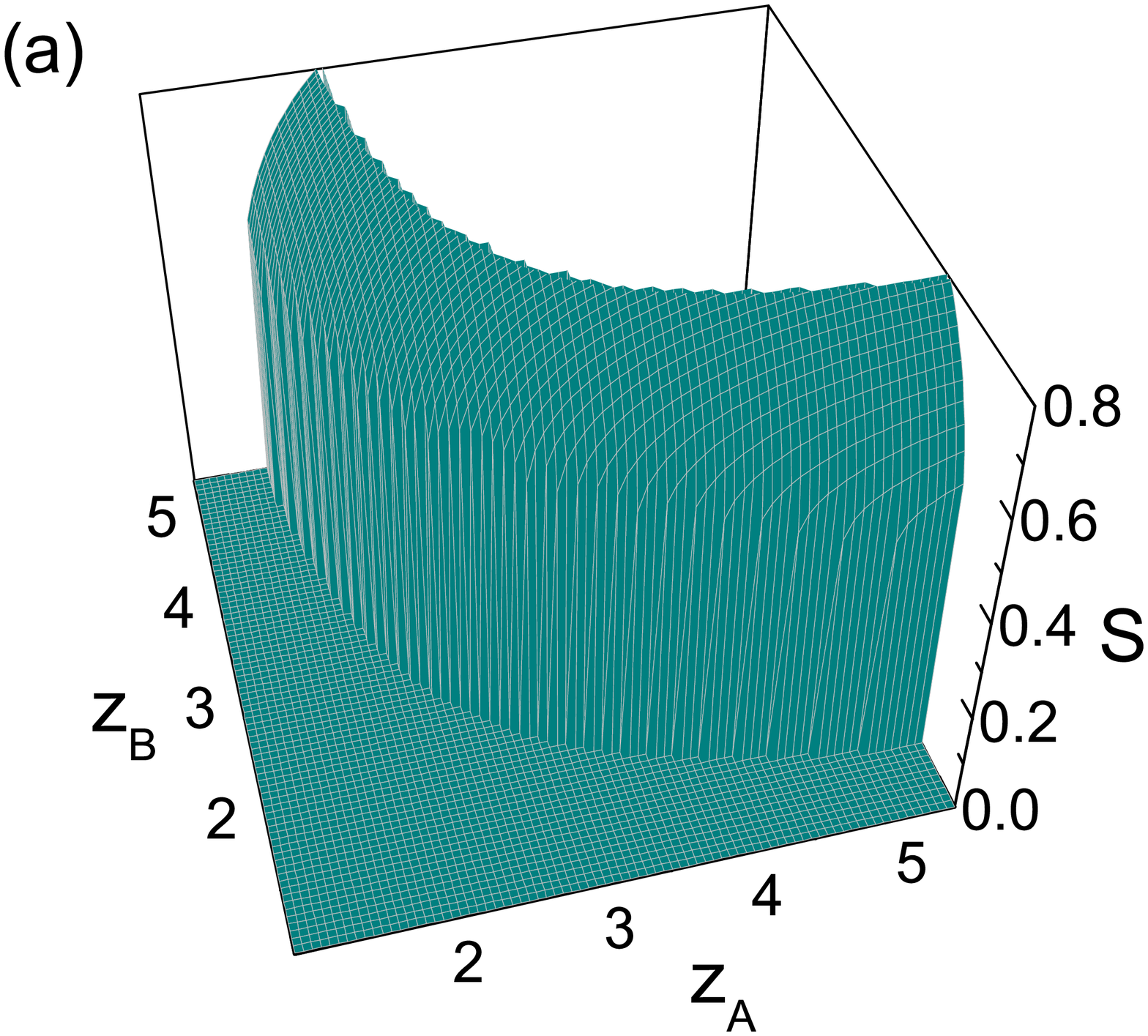}
\includegraphics[width=0.88\columnwidth]{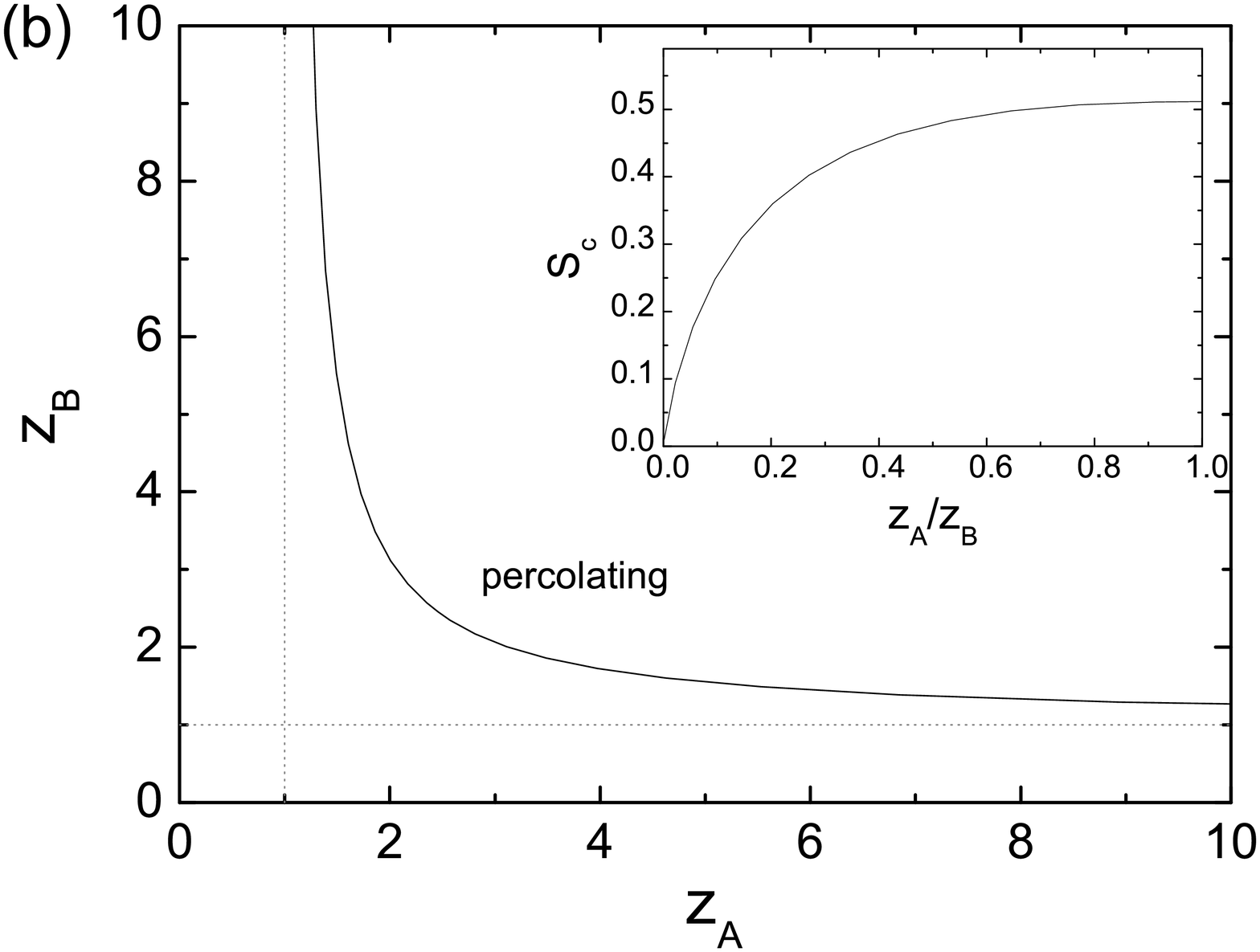}
\caption{(a) $S$ as a function of $z_{\cal A}$ and $z_{\cal B}$
for Eq.~(\ref{eq_ER}). (b) The discontinuous phase transition line
on the plane of $z_{\cal A}$ and $z_{\cal B}$. It is symmetric
about the line $z_{\cal A} = z_{\cal B}$. In the inset, the jump
size $S_c$ at the transition line is shown as a function of the
ratio $z_{\cal A} / z_{\cal B}$. } \label{fig3}
\end{figure}

For the general case $z_{\cal{A}} \neq z_{\cal{B}}$, defining \be
h(S) = S - (1-e^{-z_{\cal A}S})(1-e^{-z_{\cal B}S}),
\label{eq_8}\ee we obtain a line of discontinuous transition
points from the conditions $h(S)=0$ and $h'(S)=0$. Assuming
$h(S)=0$, the second condition can be written as \bea
    h'(S) &=& 1 - z_{\cal A} e^{-z_{\cal A}S}(1-e^{-z_{\cal B}S})
      - z_{\cal B} e^{-z_{\cal B}S}(1-e^{-z_{\cal A}S}) \nonumber \\
          &=& 1 - {{z_{\cal A}S~ e^{-z_{\cal A}S}}\over{1-e^{-z_{\cal
       A}S}}} - {{z_{\cal B}S~ e^{-z_{\cal B}S}}\over{1-e^{-z_{\cal
       B}S}}} \nonumber \\
          &=& 1 - {{x}\over{e^{x}-1}} - {{y}\over{e^{y}-1}} = 0,
\eea where  $x=z_{\cal A}S $ and $y=z_{\cal B}S$.
 This gives a one-parameter set
of solutions $y(x)$, from which $S_c$ can be obtained using
Eq.~(\ref{eq_ER}). Finally, $z_{\cal A}$ and $z_{\cal B}$ are
obtained by $z_{\cal A}=x/S$ and $z_{\cal B}=y/S$. The resulting
discontinuous phase transition line is shown in
Fig.~\ref{fig3}(b), where $z_{\cal A}=1$ and $z_{\cal B}=1$ act as
asymptotes. In the inset of Fig.~\ref{fig3}(b), the jump size
$S_c$ is shown as a function of the ratio $z_{\cal A}/z_{\cal B}$
(only $z_{\cal A}/z_{\cal B} <1$ is shown, since the transition
line is symmetric about the line $z_{\cal A}=z_{\cal B}$). The
full dependence of $S$ on $z_{\cal A}$ and $z_{\cal B}$ is shown
in Fig.~\ref{fig3}(a).

(iii) \emph{An arbitrary number of  interdependent networks} - Our
approach is easily extended to treat coupling of more than two
networks~\cite{Gao2011}. For $M$ networks, Eqs.~(\ref{eq5a}) and
(\ref{eq5}) can be simply replaced by
\bea S &=& \prod_{m=1}^M(1-{G^{m}_0(1-S'_{m})}), \nonumber \\
S'_{m}&=& \frac{1-{G^{m}_1(1-S'_{m})}}{1-{G^{m}_0(1-S'_{m})}}
~S.~\label{SS} \eea This is due to the fact that in order to be
part of the infinite percolating cluster each node must, by
definition, be connected to it via all of the $M$ networks. Order
parameters and transition points can be obtained in an analogous
way and the transition is discontinuous for any $M>1$.

(iv) \emph{Partially interdependent networks} - Assume now that
two networks are not totally interdependent. In network $\cal{A}$
only a fraction $q_{\cal{A}}$ of the nodes are dependent on a node
in network $\cal{B}$. To be on the percolating cluster, a node has
to be connected to the cluster via ${\cal A}$ links and the node
on which it depends has to be connected to the cluster via ${\cal
B}$ links. Similarly a fraction $q_{\cal B}$ of nodes in network
$\cal {B}$ depend on a node in $\cal{A}$. Note that here one node
from a network depends only on one node from the other network,
{\it i.e.}, each node can have only one dependency link, which can
be unidirectional or bidirectional~\cite{Parshani2010}. In that
case, in general we must expect that the two order parameters
$S_{\cal{A}}$ and $S_{\cal{B}}$ are different. They indicate the
chance that a randomly picked node in $\cal{A}$ (resp. $\cal{B}$)
is a member of the percolating $\cal{AB}$-cluster.

Let us first discuss the symmetric case
$q_{\cal{A}}=q_{\cal{B}}=q$. If a given node in network $\cal{A}$
does {\it not} depend on network  $\cal{B}$ (which happens with
probability $1-q$), it is a member of the percolating ${\cal
AB}$-cluster {\it iff} at least one of its neighbors in $\cal{A}$
is also a member of that cluster. On the other hand, with
probability $q$, the node does depend on a node in $\cal{B}$. In
that case, in order for it to be part of the percolating ${\cal
AB}$-cluster, its dependency partner also must be connected via
network $\cal{B}$ to at least one node in that cluster. Therefore,
the probability $S_{\cal{A}}$ can be expressed by summing the
conditional probability to be connected multiplied by the
corresponding probability to be dependent ($q$) or not ($1-q$):
\bea
    S_{\cal{A}} &=& q
       (1-G^{\cal{A}}_0(1-S'_{\cal{A}}))(1-G^{\cal{B}}_0(1-S'_{\cal{B}}))
       \nonumber \\ & & +(1-q)(1-G^{\cal{A}}_0(1-S'_{\cal{A}})) \nonumber
       \\ &=&
       (1-G^{\cal{A}}_0(1-S'_{\cal{A}}))(1-qG^{\cal{B}}_0(1-S'_{\cal{B}})).~\label{eq12}
\eea Similarly, \be
    S'_{\cal{A}} =
    (1-G^{\cal{A}}_1(1-S'_{\cal{A}}))(1-qG^{\cal{B}}_0(1-S'_{\cal{B}})).~\label{eq13}
\ee By symmetry, another pair of conditions exists for network
$\cal{B}$: \bea
    S_{\cal{B}} &=&
    (1-G^{\cal{B}}_0(1-S'_{\cal{B}}))(1-qG^{\cal{A}}_0(1-S'_{\cal{A}})),
    \nonumber \\ S'_{\cal{B}} &=&
    (1-G^{\cal{B}}_1(1-S'_{\cal{B}}))(1-qG^{\cal{A}}_0(1-S'_{\cal{A}})).
\eea More generally, if $q_{\cal{A}} \neq q_{\cal{B}}$, \bea
   S_{\cal{A}} &=&
   (1-G^{\cal{A}}_0(1-S'_{\cal{A}}))(1-q_{\cal{A}}G^{\cal{B}}_0(1-S'_{\cal{B}})),
   \nonumber \\ S'_{\cal{A}} &=&
   (1-G^{\cal{A}}_1(1-S'_{\cal{A}}))(1-q_{\cal{A}}G^{\cal{B}}_0(1-S'_{\cal{B}})),
   \nonumber \\ S_{\cal{B}} &=&
   (1-G^{\cal{B}}_0(1-S'_{\cal{B}}))(1-q_{\cal{B}}G^{\cal{A}}_0(1-S'_{\cal{A}})),
   \nonumber \\ S'_{\cal{B}} &=&
   (1-G^{\cal{B}}_1(1-S'_{\cal{B}}))(1-q_{\cal{B}}G^{\cal{A}}_0(1-S'_{\cal{A}})).
\eea Again, these four coupled equations involve nothing more
complicated than (weighted) products of the network-specific
fundamental factors from Eqs.~(\ref {eq2}) and (\ref {eq3}).
Together, they give the behavior of the order parameters
$S_{\cal{A}}$ and $S_{\cal{B}}$. These expressions are completely
equivalent to the more complicated results in~\cite{Parshani2010}.
If $q_{\cal{A}} = q_{\cal{B}} =q =1$, these equations reduce to
Eqs.~(\ref{eq5a}) and (\ref{eq5}).

If $q_{\cal{A}} = q_{\cal{B}} =q \neq 1$ and if the two networks
have the same degree distribution, $S_{\cal{A}} = S_{\cal{B}} =
S$.  For example, if two ER networks having the same mean degree
$z$ are coupled with the dependency probability $q \neq 1$, the
solution is simply \be
    S = (1-e^{-zS})(1-qe^{-zS}). \label{eq10}
\ee The solution $S$ can now be expressed as a function of both
$z$ and $q$. When $q=0$, $S(z)$ shows a continuous transition,
since now both networks are fully independent. On the other hand,
when $q=1$, the transition is discontinuous. A crossover from
\emph{second order} behavior to \emph{first order } behavior at
the tricritical point $q_c$ is observed as $q$ is increased from 0
to 1. In analogy to Eq.~(\ref{eq_8}), we now define \be
    h(S) = S - (1 - e^{-zS})(1 - q e^{-zS}), \label{eq14}
\ee The tricritical point is found by demanding $S=h(S) = h'(S)=
h''(S) =0$, which gives $q_c = 1/3$ and $z_c = 3/2$
\cite{Parshani2010}.

\begin{figure}
\center\includegraphics[width=0.8\columnwidth]{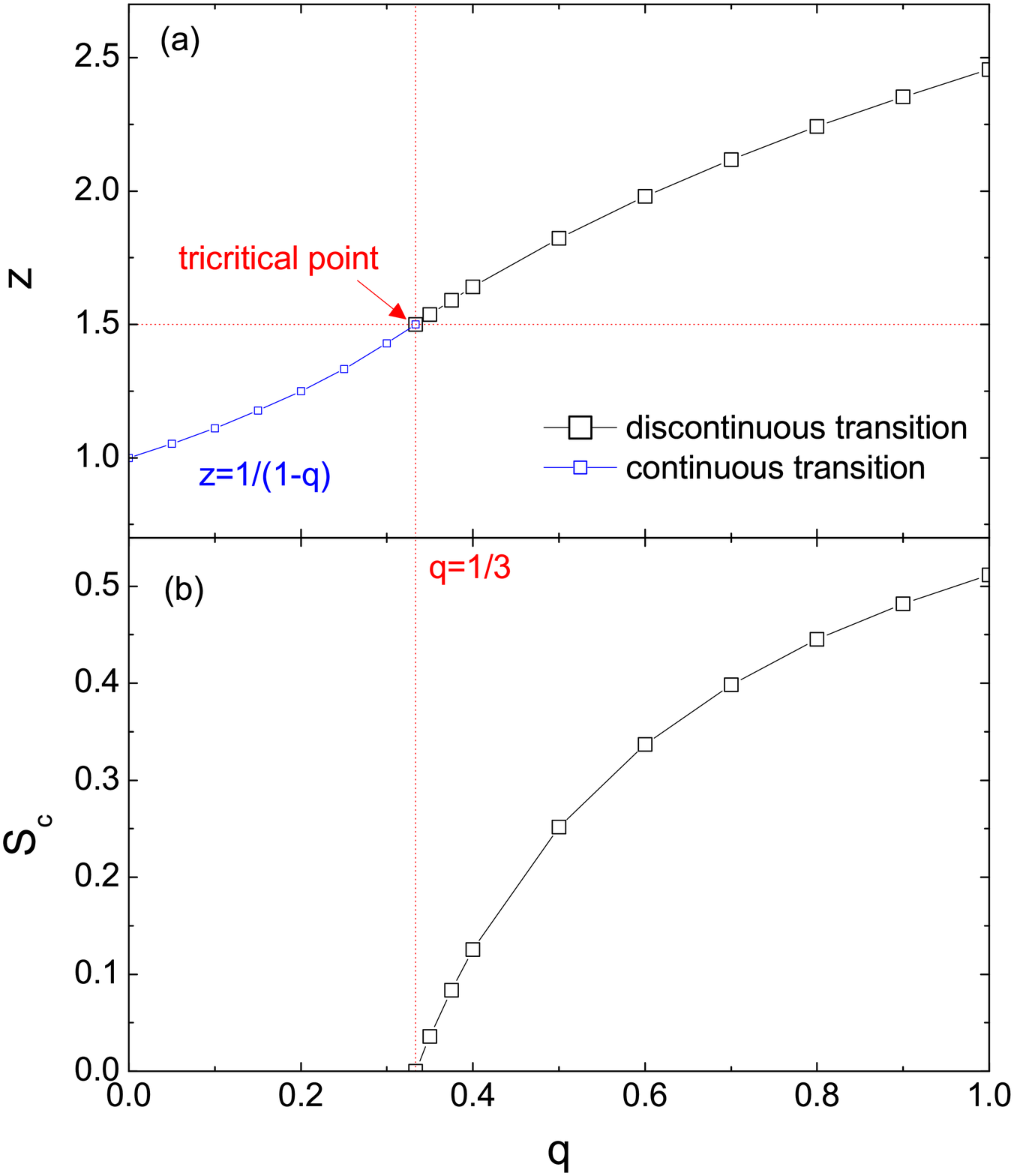}
\caption{(Color online) (a) Percolation transition points for two
partially coupled ER networks with the same mean degree $z$ at
different values of the  dependency $q$. (b) The jump size $S_c$
at the transition point. The first order phase transition line ($q
> 1/3$) meets the second order phase transition line ($q < 1/3$)
at $q = 1/3$. The red dotted line indicates the tricritical point
$(q_c, z_c) = (1/3, 3/2)$.} \label{fig4}
\end{figure}

For $q>q_c$, we can find the first order transition point by
considering $h(S)=0$ and $h'(S)=0$. The subsequent analysis is
straightforward and follows closely our previous method. Its
results are displayed in Fig.~\ref{fig4}.

\begin{figure}[t]
\center\includegraphics[width=0.7\columnwidth]{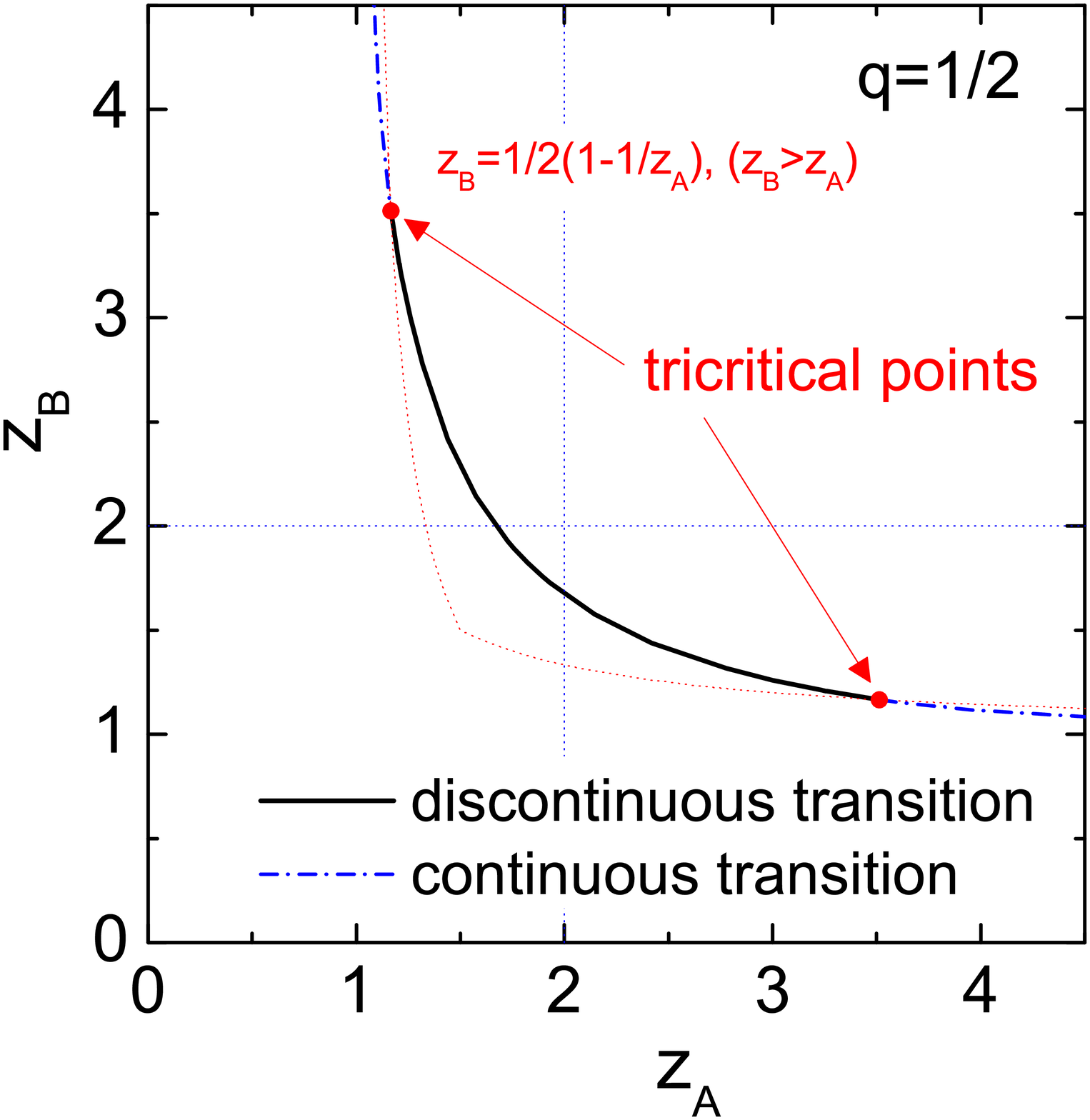}
\caption{(Color online) Phase transition line at $q=1/2$ for
different $z_{\cal{A}}$ and $z_{\cal{B}}$. The black solid line
indicates the first order transition line and the blue dash-dot
line represents the second order transitions. They meet two
tricritical points (red bullets). The tricritical points move
along the red dotted line as $q$ changes. When $q=1/3$, the
tricritical points meet at $z_{\cal{A}}=z_{\cal{B}}=3/2$.}
\label{fig5}
\end{figure}

Let  us briefly discuss the case $q_{\cal{A}} = q_{\cal{B}} = q$
and $z_{\cal{A}}\neq z_{\cal{B}}$ of ER networks, which was not
treated before. In this case, also $S_{\cal{A}}\neq S_{\cal{B}}$,
and we have to solve the coupled equations \bea
&h_{\cal{A}}(S_{\cal{A}}) = S_{\cal{A}} -
         (1-e^{-z_{\cal{A}}S_{\cal{A}}})(1-qe^{-z_{\cal{B}}S_{\cal{B}}}) =0,&\nonumber \\
&    h_{\cal{B}}(S_{\cal{B}}) = S_{\cal{B}} -
         (1-e^{-z_{\cal{B}}S_{\cal{B}}})(1-qe^{-z_{\cal{A}}S_{\cal{A}}})
         =0.&
\quad \label{eq16} \eea The percolation transition is obtained by
imposing in addition
$h'_{\cal{A}}(S_{\cal{A}})=h'_{\cal{B}}(S_{\cal{B}})=0$. Defining
again $x=z_{\cal{A}} S_{\cal{A}}$ and $y=z_{\cal{B}} S_{\cal{B}}$,
we find \be
   \left(1-{{x}\over{e^{x}-1}} \right) \left(1-{{y}\over{e^{y}-1}}
   \right) = {{qx}\over{e^{x}-q}} \times {{qy}\over{e^{y}-q}}~~.
\ee For given $q$ this is solved graphically (Fig.~\ref{fig5}).
The values of $S_{\cal{A}}$ and $S_{\cal{B}}$ at the transition
point are then obtained from Eq.~(\ref{eq16}). If they vanish, the
transition is continuous, otherwise it is discontinuous. Results
are shown in Fig.~\ref{fig5} for $q=1/2$. If $q$ decreases, the
two tricritical points move together. They coalesce at
$z_{\cal{A}}=z_{\cal{B}}=3/2$ when $q\to 1/3$.

(v) \emph{Dependency networks} -  A dependency
network~\cite{Parshani2011,Bashan2011} is a \emph{single} network
that contains two types of links, connectivity links and
dependency links.  In the simplest case each node in the network
depends on one other node in that network and all dependencies are
mutual. In order for a node to be connected to the infinite
self-sustaining cluster both it and its partner must be part of
that cluster.  For random networks this leads immediately to
Eqs.~(\ref{eq5a}) and (\ref{eq5}) where the subscripts ${\cal A}$
and ${\cal B}$ are dropped since there is only one network.  In
the case that only a fraction $q$ of nodes have dependency links
we get Eqs.~(\ref{eq12}) and (\ref{eq13}), again dropping the
subscripts labelling the networks. Again the type of transition
depends on the value of $q$, with a tricritical point separating
the two regimes.

In the most general case where a node has $m$ dependency links
with probability $p(m)$ the resulting equations are \bea
&& S=\sum_{m=0}^{N-1} p(m)(1 - G_0(1-S'))^{m+1}, \nonumber \\
&& S'= (1- G_1(1-S'))\sum_{m=0}^{N-1} p(m)(1- G_0(1-S'))^{m}.
\nonumber \eea Again the precise behavior reflects  the extent to
which the network is dependent, with a low dependency regime
exhibiting a continuous transition in the universality class of
ordinary percolation and a high dependency regime exhibiting a
first order transition, with crossover controlled by a tricritical
point. Similar arguments can  be used to derive the general case
for interdependent networks where a single node has dependencies
to $m$ other networks with probability $p(m)$. In that case the
equations are more complicated because each network's structure
may be different and they will have different order parameters,
but the arguments used to derive the equations are precisely the
same. The above does not apply to networks with directed
(non-mutual) dependencies, for which different arguments apply.

In summary, we consider percolation on various interdependent or
dependency networks, pointing out the close analogy to epidemic
spreading on single networks without these dependencies. Our
arguments are much more straightforward, both conceptually and
mathematically, than previous ones built on cascades of failures.
We should however stress that they apply, like those of
\cite{Buldyrev2010,Parshani2010,Parshani2011,Gao2011,Bashan2011}
only to random locally treelike graphs. For interdependent
networks that are correlated with each other \cite{Parshani2010a}
or that are spatially embedded \cite{Son2011,Son2011a} the
transition is in general not first order, and the interdependency
can make the transition even less sharp than in ordinary
percolation \cite{Son2011,Son2011a}. For these more realistic
cases no analytical theory is yet available.

\end{document}